\documentclass[12pt,a4paper]{article}

\usepackage{graphicx}
\usepackage{amsfonts}
\usepackage{amsmath}
\usepackage{hyperref}
\usepackage{psfrag,subfigure}

\usepackage[numbers,square,comma, compress]{natbib} 

\csname @addtoreset\endcsname{equation}{section}

\usepackage[lmargin=1.7 cm, rmargin=1.7 cm, tmargin=3.0cm, bmargin=3.0cm]{geometry}

\usepackage{float}

\title{
\bf{$\boldsymbol{\mathcal{N}=2}$ string amplitudes and the Omega background}}
\author{Ioannis Florakis\footnote{{\tt florakis@mppmu.mpg.de}, Speaker}~ and Ahmad Zein Assi\footnote{{\tt zeinassi@cern.ch}, Speaker}
}
\date{}

\begin{document}

\maketitle

\begin{center}
\renewcommand{\thefootnote}{\fnsymbol{footnote}}\vspace{-0.5cm}
${}^{\footnotemark[1]}$ Max-Planck-Institut f\"{u}r Physik,
	    Werner-Heisenberg-Institut,
              80805 M\"{u}nchen, Germany\\[0.2cm]
${}^{\footnotemark[2]}$ Department of Physics, CERN - Theory Division, CH-1211 Geneva 23, Switzerland\\[0.2cm]

\end{center}

\begin{abstract}
 We review recent work \cite{Antoniadis:2013bja,Antoniadis:2013mna} on obtaining a realisation of the $\Omega$-background in terms of a special series of higher-derivative generalised F-terms in the effective $\mathcal{N}=2$ supergravity action. We discuss the motivation behind the identification of these couplings, their relation to the Nekrasov partition function and connect them to a worldsheet approach towards the refined topological string.
\end{abstract}

\vspace{2mm}
\begin{center}
	\emph{Presented by the authors at the Corfu Summer Institute 2013 ``Workshop on Noncommutative Field Theory and Gravity", September 8-15, 2013\\ Corfu, Greece}
\end{center}




\section{Introduction}

During the last decades, the interplay between string theory, the topological string and supersymmetric gauge theories has led to fascinating developments in mathematics as well as in physics. For instance, from the mathematical point of view, the observables in topological string theory can usually be understood in terms of topological invariants which acquire a physical meaning in string theory. Indeed, considered as a twisted version of the latter, the topological string captures properties of string theory that depend on its BPS sector only and can thus be viewed as a subsector of string theory.

In order to make this connection more precise, consider a Calabi-Yau compactification of type II string theory and perform the topological twist on the underlying superconformal algebra (SCA), \emph{i.e.} shifting the energy-momentum tensor with the $U(1)$ current as $T\rightarrow T+\tfrac{1}{2}\partial J$ (and similarly for the right-movers). The resulting theory can be shown to be topological \cite{Witten:1988xj}. On the other hand, the twisting of the SCA can be implemented at genus $g$ in type II through a deformation of its $\sigma$-model by the operator $\frac{i\sqrt{3}}{2}\int_{\Sigma_g} R^{(2)}\,H$, see \cite{Antoniadis:1996qg}, where $R^{(2)}$ is the two-dimensional scalar curvature and $H$ is the free scalar bosonising $J$, \emph{i.e} $J=i\sqrt{3}\partial H$. By choosing a metric such that $R^{(2)}=-\sum_{i=1}^{2g-2} \delta^{(2)}(x-x_i)$, it is easy to see that the partition function in the twisted, topological theory can be calculated in the untwisted, type II theory as an amplitude involving  $2g-2$ insertions of vertex operators whose internal part is given by $e^{-\frac{i\sqrt{3}}{2}H}$. The latter is identified as the internal part of the graviphoton vertex operator. Hence, it turns out that the genus $g$ topological string partition function $\mathcal F_g$ computes the coupling coefficients of a class of higher-derivative F-terms in the string effective action at genus $g$ \cite{Antoniadis:1993ze}, given by
\begin{equation}\label{GraviphotonCoupling}
 \mathcal F_g(\varphi)\, R^2_{(-)} F^{2g-2}_{G,(-)}\,,
\end{equation}
with $R_{(-)}$ being the anti-self-dual Riemann tensor and $F_{G,(-)}$ the anti-self-dual part of the graviphoton field strength\footnote{The necessity of the two Riemann tensor insertions can be seen to arise either by supersymmetry arguments or, at the technical level, in order to soak up fermionic zero-modes in the string amplitude.}. They originate from the supersymmetric F-term involving the Weyl multiplet $W$,
\begin{equation}\label{FtermW}
 \int d^4\theta\, \mathcal F_g(X) W^{2g}\,,
\end{equation}
where $X$ are chiral vector multiplets with scalar component $\varphi$.

From the low-energy perspective, one of the uses of topological string theory is to geometrically engineer supersymmetric gauge theories in four dimensions \cite{Gopakumar:1998ii}. For example, consider the local $\mathbb{P}^1\times \mathbb{P}^1$ model arising from a type IIA compactification on an elliptically fibered $K3$ over $\mathbb{P}^1$, which is used to engineer the $\mathcal N=2$, $SU(2)$ gauge theory of Seiberg and Witten. The genus zero partition function of this model, $\mathcal F_0$, coincides, in the field theory limit, with the prepotential of the supersymmetric gauge theory. In addition, $\mathcal F_{g>1}$ can be interpreted as gravitational corrections to the Seiberg-Witten prepotential encoded in the F-terms \eqref{FtermW}. In fact, one can understand these corrections gauge-theoretically through the $\mathcal N=2$ gauge theory in the $\Omega$-background \cite{Losev:1997bz, Moore:1997dj}. More specifically, it was shown that the Seiberg-Witten solution can be obtained by a direct path integral evaluation if one introduces an $S^1$-fibration over $\mathbb R^4$ such that going around the circle is accompanied by rotations in the two $\mathbb R^2$-planes of space-time and an $R$-symmetry rotation parametrised by $\epsilon_{1,2}$ and $\frac{\epsilon_1+\epsilon_2}{2}$ respectively. The latter is crucial to preserve a fraction of supersymmetry and evaluate the path integral using localisation. Similarly to the Seiberg-Witten case, the $\Omega$-deformed prepotential receives a one-loop correction and a series of instanton contributions only:
\begin{equation}\label{NekPartitonGeneral}
 \mathcal F^{\textrm{Nek}}(\epsilon_{1},\epsilon_2)=\mathcal F^{\textrm{Nek}}_{\textrm{1-loop}}(\epsilon_{1},\epsilon_2)+\mathcal F^{\textrm{Nek}}_{\textrm{inst}}(\epsilon_{1},\epsilon_2)\,,
\end{equation}
and its logarithm is known as the Nekrasov partition function. It was shown that in the limit $\epsilon_{1,2}\rightarrow0$, the leading term in \eqref{NekPartitonGeneral} matches the Seiberg-Witten solution and, for $\epsilon_1=-\epsilon_2$, \eqref{NekPartitonGeneral} is the field theory limit of the topological string partition function, with the topological string coupling being identified with the parameter $\epsilon_1=-\epsilon_2$. This opened up the possibility of `refining' the topological string through a one-parameter extension of its partition function such that its field theory limit matches the gauge theory partition function in a general $\Omega$-background. From the worldsheet point of view of the untwisted theory, this would correspond in \eqref{GraviphotonCoupling} to additional insertions of self-dual field strenghts. One of the main difficulties in achieving this is to simultaneously realise the $R$-symmetry rotation at the worldsheet level using physical operators. Even though the existence of such objects is \emph{a priori} not guaranteed, in what follows we show that this is indeed the case by reviewing the results of \cite{Antoniadis:2013bja,Antoniadis:2013mna}.

To date, most descriptions of the refinement do not follow a worldsheet approach. For instance, the refined topological partition function can be defined as a BPS index in M-theory counting spinning M2-branes \cite{Dijkgraaf:2006um} and evaluated using the refined topological vertex techniques \cite{Awata:2005fa, Iqbal:2007ii}. Similarly to the instanton calculus in gauge theory, the latter is applicable in the asymptotic region of the moduli space only. A worldsheet realisation of the refined topological string is therefore essential in order to define it at any point in the moduli space. In this perspective, we consider a  (compact) Calabi-Yau compactification of type II theory admitting a heterotic dual and identify the additional self-dual insertions. Even though we consider the compact Calabi-Yau case\footnote{It is widely believed that the refined topological string exists only for non-compact Calabi-Yau manifolds.}, it is worth mentioning that our primary goal is to identify the full $\Omega$-background in a physical string theory setup. Furthermore, we expect that the couplings we calculate define a BPS index in type II theory once the heterotic/type II duality map is applied and the appropriate non-compact/local limit is taken.

In this note, we review the class of generalised F-terms extending \eqref{FtermW} introduced in \cite{Antoniadis:2013bja} and motivate the identification of the additional insertions. Then, in Section \ref{DeformedTopoAmp}, we discuss the evaluation of these couplings and their connection to the gauge theory partition function. Section \ref{Sec:Conclusions} highlights some of the open questions and directions for future investigation.


\section[\texorpdfstring{$\boldsymbol{\mathcal{N}=2}$ Higher Derivative Couplings}{N=2 Higher Derivative Couplings}]{\texorpdfstring{$\boldsymbol{\mathcal{N}=2}$ Higher Derivative Couplings}{N=2 Higher Derivative Couplings}}\label{Motivation}

Consider a five-dimensional $\mathcal{N}=2$ gauge theory on $\mathbb{R}^4\times S^1$ and define its partition function as the trace over the four-dimensional Hilbert space of the theory\cite{Nekrasov:2002qd}
\begin{align}\label{NekrasovTrace}
	Z_{\textrm{Nek}}(\epsilon_{-},\epsilon_{+}) = \textrm{Tr}\,(-)^F\,e^{-2\epsilon_{-}J_{-}}e^{-2\epsilon_{+}(J_{+}+J_R)}\,e^{-\beta H}\,,
\end{align}
where $F$ is the fermion number, $J_{\pm}$ are the Cartan generators of the $SU(2)_{-}\times SU(2)_{+}$ subgroups the Euclidean rotation group $SO(4)$ of $\mathbb{R}^4$, $J_R$ is the Cartan generator of the $SU(2)$ $R$-symmetry and the `inverse temperature' $\beta$ is identified with the radius of $S^1$. This trace may be equivalently described as a vacuum amplitude on a deformed Melvin-like background, $ds^2=(dX^\mu+\Omega^\mu dX)^2+(dX)^2$, with $X$ being the $S^1$ coordinate. This space, known as the $\Omega$-background, is a fibration of $\mathbb{R}^4$ over $S^1$, with the $\mathbb{R}^4$-fibre being twisted by isometry operations encoded into two constant rotation parameters $\epsilon_{1,2}$, through $d\Omega^\mu =\epsilon_1 dX^1\wedge dX^2+\epsilon_2 dX^3\wedge dX^4$, with the identification $\epsilon_{\pm}=\frac{1}{2}(\epsilon_1\pm\epsilon_2)$.

In order to motivate the definition of the generalised F-term couplings of \cite{Antoniadis:2013bja}, it is instructive to obtain a deeper understanding of the relation between the string amplitudes $\langle R^2 F_G^{2g-2}\rangle$ and the Nekrasov partition function for $\epsilon_{+}=0$. The connections are summarised in the table below.
\begin{align}\nonumber
	\begin{array}{c | c | c}
		\textrm{Twisted theory} & \textrm{Physical Couplings} & \textrm{Gauge Theory}\\ \hline\hline
		\textrm{Top. String}~ \mathcal F_{g} & R^2_{(-)} F_{G,(-)}^{2g-2} & Z_{\textrm{Nek}}(\epsilon_{-},0)\\ \hline
		\textrm{Refined Top. String}~ \mathcal F_{g,n} & R^2_{(-)} F_{G,(-)}^{2g-2} F_{{X}^\dagger,(+)}^{2n} & Z_{\textrm{Nek}}(\epsilon_-,\epsilon_+)\\
	\end{array}
\end{align}
Consider the scattering of two gravitons and $2g-2$ graviphotons in type II string theory on a Calabi-Yau threefold. This amplitude first appears at genus $g$ and, being topological, receives no perturbative or non-perturbative corrections, consistently with $\mathcal{N}=2$ super-renormalisation theorems. Let us focus on the perturbative part of the $\mathcal F_g$. Using type II-heterotic duality, it may be obtained by evaluating a one-loop amplitude of the same coupling in the dual heterotic theory on $K3\times T^2$ in the weak coupling limit. Complexifying the space-time directions as $Z^1,\bar{Z}^1,Z^2,\bar{Z}^2$, we write the relevant vertex operators in a kinematic configuration consistent with the projection on the anti-self-dual components of the corresponding tensors:
\begin{align}\label{GraviphotonVertexOps}
	\begin{split}
	V_{\textrm{graviton}}(p_1)&= (\partial Z^2-ip_1\chi^1\chi^2)\bar\partial Z^2\,e^{ip_1 Z^1}\,,\\
	V_{\textrm{graviton}}(\bar p_1)&= (\partial \bar Z^2-i\bar p_1\bar\chi^1\bar\chi^2)\bar\partial \bar Z^2\,e^{i\bar p_1 \bar Z^1}\,,\\
	V_{\textrm{graviph.}}(p_1)&= (\partial X-ip_1\chi^1 \Psi)\bar\partial Z^2\,e^{ip_1 Z^1}\,\,\,\,,\\
	V_{\textrm{graviph.}}(\bar p_1)&= (\partial X-i\bar p_1\bar\chi^1 \Psi)\bar\partial \bar Z^2\,e^{i\bar p_1 \bar Z^1}\,\,\,\,.\\
	\end{split}~
\end{align}
Let us briefly outline the non-technical aspects of the argument. The effect of the graviton insertions is to soak up the zero modes of the worldsheet fermions in the space-time directions $\chi^\mu$. The compexified worldsheet fermions in the $T^2$ direction $\Psi$ do not contract and their one-loop contribution cancels against that of the superghost. Working in the orbifold limit of $K3\sim T^4/\mathbb{Z}_N$, the twisted one-loop contribution of the $K3$ fermions, cancels against the (left-moving) twisted bosonic $K3$ contribution and the full correlator has the form of a BPS amplitude involving a non-trivial bosonic correlator, weighted by the partition function of the theory which produces the elliptic genus. Since only the complexified $T^2$ coordinate $X$ appears in the correlator, but never its complex conjugate, it leads to no contractions but, rather, contributes only through left-moving zero modes $\langle\partial X\rangle$. In the Hamiltonian representation of the lattice, this is replaced by left-moving lattice momentum insertions $\langle\partial X\rangle\rightarrow \sqrt{T_2 U_2}\,\tau_2 P_L= [m_2-Um_1+T(n_2+Un_1)]\tau_2$, with $T,U$ being the K\"ahler and complex structure moduli of $T^2$, respectively, and $\tau=\tau_1+i\tau_2$ being the complex structure of the worldsheet torus. The bosonic correlator then becomes
\begin{align}
	(p_1)^{g-1} (\bar{p}_1)^{g-1} \,\langle\,\prod_{i}^{g-1} \partial X (Z^1\bar\partial Z^2)(x_i) \prod_{j}^{g-1} \partial X (\bar{Z}^1\bar\partial \bar{Z}^2) (y_j)\,\rangle\,,
\end{align}
where we suppress the integrals over the positions $x_i,y_i$ on the worldsheet torus. Multiplying this correlator by $\epsilon_{-}^{2g-2}$, stripping off its kinematical factors and summing over $g$ expresses the bosonic correlator in terms of a generating function
\begin{align}\label{BosCorr}
	G_{\textrm{boson}}(\epsilon_{-})= \,\Bigr\langle \exp\Bigr[-\epsilon_{-}\int d^2 z~ \partial X(Z^1\bar\partial Z^2+\bar{Z}^2\bar\partial \bar Z^1)\Bigr]\Bigr\rangle \equiv \langle e^{-\tilde\epsilon_{-}J_{-}}\rangle\,,
\end{align}
where $\tilde{\epsilon}_{-}=\epsilon_{-}\tau_2 P_L \sqrt{T_2 U_2}$ and $J_{-}=\int(Z^1\bar\partial Z^2+\bar{Z}^2\bar\partial Z^1)$ is the string generator of $SU(2)_{-}$ rotations, defining a map from the worldsheet to target space. In the field theory limit $\tau_2\rightarrow\infty$, the dependence on $\tau$ disappears and $J_{-}$ counts the $SU(2)_{-}$ charges of the BPS states becoming massless as $P_L\rightarrow 0$. Indeed, at the self-dual $T=U$ point in moduli space, the gauge group arising from $T^2$ acquires an $SU(2)$ enhancement and the modular integral over $\tau$ is dominated by the field theory contribution. Thus, the bosonic correlator \eqref{BosCorr} only contributes through its zero mode:
\begin{align}
	G_{\textrm{boson}}(\epsilon_{-}) \sim \frac{(2\pi\tilde\epsilon_{-})^2\bar\eta^6}{\bar\vartheta_1(\tilde\epsilon_{-})^2}\,e^{-\frac{\pi\tilde\epsilon_{-}^2}{\tau_2}} ~\rightarrow~ \frac{\pi^2\epsilon_{-}^2}{\sin^2\tilde\epsilon_{-}} + \mathcal{O}(e^{-\pi\tau_2})\,,
\end{align}
and one obtains the result in the form of a one-loop Schwinger integral
\begin{align}
	\mathcal F(\epsilon_{-}) \sim \int_0^\infty\frac{dt}{t}\,\Bigr(\frac{\pi\epsilon_{-}}{\sin\epsilon_{-}t}\Bigr)^2\,e^{-\mu t}\,,
\end{align}
with $\mu\sim T-U$ being the BPS mass parameter. Therefore, the physical picture is that the scattering of graviphotons, upon exponentiation as in \eqref{BosCorr}, generates a deformed background\footnote{More details on the relation between Euclidean traces with chemical potentials \eqref{NekrasovTrace} and deformed one-loop string vacuum energies may be found in \cite{Florakis:2010ty} and in references therein.} which precisely mimics the $\Omega$-background in \eqref{NekrasovTrace} for $\epsilon_{+}=0$.

Inspired by the Gopakumar-Vafa reformulation of the topological string, the additional insertions should correspond to self-dual field strength vertices $F_{X^\dagger,(+)}^{2n}$ in the anti-chiral vector multiplet $X^\dagger$ such that they couple to the spins of the $SU(2)_{+}$ component of $SO(4)$, as required by \eqref{NekrasovTrace}. The known way of introducing self-dual field strength insertions to the graviphoton couplings \eqref{GraviphotonCoupling} is by means of a generalised F-term\begin{align}\label{GenFterms}
	\int d^4\theta~\tilde{\mathcal F}_{g,n}(X) W^{2g}\Upsilon^{2n} = \mathcal F_{g,n}(\varphi,\varphi^\dagger)\,R^2_{(-)}F_{G,(-)}^{2g-2}F_{X^\dagger,(+)}^{2n}+\ldots\,,
\end{align}
with $\Upsilon=\Pi\circ f(X,X^\dagger)$ being an $\mathcal{N}=2$ chiral projection\footnote{The reason why this is not a true F-term is that the projection operator $\Pi=(\epsilon_{ij}\bar{D}^i\bar\sigma\bar{D}^j)^2$ may be pulled in front of $\tilde{F}_{g,n}(X)W^{2g}$ in order to modify the Grassmann integration measure and recast the coupling in the form of a D-term.}  of an arbitrary function $f(X,X^\dagger)$ of both chiral and anti-chiral vector multiplets. We now exploit the relation between the deformed (exponentiated) background \eqref{BosCorr} and the trace representation of the gauge theory partition function in the $\Omega$ background \eqref{NekrasovTrace} in order to motivate the specific choice of species $X^\dagger$ of the self-dual vector insertions. Indeed, for the bosonic fields, we would like to obtain a coupling of $\epsilon_{-}$ to $J_{-}$ and $\epsilon_{+}$ to the analogous $SU(2)_{+}$ generator. \emph{A priori}, there are two possible choices:
\begin{align}
	J_{+} = \int (Z^1\bar\partial \bar{Z}^2+Z^2\bar\partial\bar Z^1) \qquad\textrm{or}\qquad \check{J}_{+}=\int(Z^1\partial \bar{Z}^2+Z^2\partial\bar Z^1)\,,
\end{align}
depending on whether the $SU(2)_{+}$ current arises from the left- or the right- moving part of the self-dual vertex operator insertions. In the corresponding correlators, $J_{+}$ is accompanied by $\tilde\epsilon_{+}=\langle\partial X\rangle\epsilon_{+}$, whereas $\check{J}_{+}$ is accompanied by $\check\epsilon_{+}=\langle\bar\partial X\rangle \epsilon_{+}$. Notice that we do not allow $\partial\bar X$ or $\bar\partial\bar X$ insertions, since that would spoil the requirement of exact solvability of our amplitude. Indeed, the factorisation of the full correlator into its bosonic and fermionic parts and the exact evaluation of the exponentiated background as a Gaussian path integral, yielding a correlator with good modular properties, is only possible in the absence of contractions in the $T^2$ coordinates.

Consider first the $J_{+}$ current. Since it is accompanied by a left-moving $\partial X$, the corresponding vertex operator is of the form $(\partial X-ip_\mu\chi^\mu\Psi)\bar\partial Z^\mu e^{ip\cdot Z}$. Hence, the left-moving part of the vertex is very similar to the graviphoton one and the resulting amplitude is BPS at the one-loop level in the heterotic theory. One can show that the self-dual projection of this vertex operator sits in the anti-chiral dilaton multiplet $\bar{S}$. This amplitude was first computed in \cite{Morales:1996bp} and its type II interpretation was further studied in \cite{Antoniadis:2010iq}. However, as was the case with the graviphoton vertices \eqref{GraviphotonVertexOps}, there is no non-trivial contribution from the worldsheet fermions. As a result, it is impossible to correctly mimic the $R$-symmetry twist and the resulting coupling function, in the field theory limit, fails to reproduce the correct (perturbative) gauge theory partition function.

On the other hand, the situation is drastically different if one considers instead the $\check{J}_{+}$ current. Since this is left-moving, it is accompanied by the right moving $\bar\partial X$ current and the structure of the self-dual vertex operator insertions takes the form
\begin{align}\label{Tvertices}
	\begin{split}
	V_{(+)}(p_1)& = (\partial \bar{Z}^2-ip_1\chi^1\bar\chi^2)\bar\partial X\,e^{ip_1 Z^1}\,,\\
	V_{(+)}(\bar p_1)& = (\partial Z^2-i\bar{p}_1\bar\chi^1\chi^2)\bar\partial X\,e^{i\bar{p}_1 \bar{Z}^1}\,.\\
	\end{split}
\end{align}
It can be checked that the corresponding bosonic correlator indeed exponentiates to $G_\textrm{bos}(\epsilon_{-},\epsilon_{+})=\langle\exp[-\tilde\epsilon_{-}J_{-}-\check\epsilon_{+}\check{J}_{+}]\rangle$, as required. However, notice that the self-dual vertex operators may also contribute through their fermionic bilinears parts, $\chi^1\bar\chi^2$ and $\bar\chi^1\chi^2$. This opens up the possibility of generating the correct $R$-symmetry twist, but only in an effective way at the string level since the internal $K3\times T^2$ space is compact. Indeed, the operator rotating the supercharges is written in terms of the complex $K3$ fermions $\chi^4,\chi^5$ and, in a compact space, rotations in the $R$-symmetry space may only be defined for quantised angles (the corresponding bosonic operator is not a well-defined conformal field). In the present case, it turns out that there is a way to obtain an effective $R$-symmetry twist after summing over the spin structures. In order to show this, suppose that $2m$ self-dual vertices at positions $u_i, u'_i$ contribute their fermion bilinear piece:
\begin{align}\nonumber
	\Bigr\langle\chi^1\chi^2(x)\bar\chi^1\bar\chi^2(y)\prod_i^m \chi^1\bar\chi^2(u_i)\prod_j^m\bar\chi^1\chi^2(u'_i)\Bigr\rangle \sim \vartheta_s(x-y+u-u')\vartheta_s(x-y-u+u')\vartheta_{s,h}(0)\vartheta_{s,-h}(0)\,,
\end{align}
where $u,u'$ is a short-hand notation for $\sum_i u_i$ and $\sum_i u_i'$, respectively, and $x,y$ are the positions of the graviton vertices. The r.h.s.\footnote{For simplicity, we do not display spin-structure independent prime forms.} displays the one-loop Jacobi $\vartheta$-functions at spin structure~$s$. The first two correspond to the space-time directions, while the last two correspond to the $K3$ fermionic directions, twisted by the orbifold action $h$. Summing over the spin structures with the conventional Riemann summation identity, the r.h.s. becomes a product of correlators in the odd spin structure:
\begin{align}\nonumber
	\vartheta_1(x-y)\vartheta_1(x-y)\vartheta_{1,h}(u-u')\vartheta_{1,-h}(u-u')\sim \langle\chi^1\chi^2(x)\bar\chi^1\bar\chi^2(y)\rangle\Bigr\langle\prod_i^m \chi^4\chi^5(u_i)\bar\chi^4\bar\chi^5(u'_i)\Bigr\rangle_h\,.
\end{align}
The fermions in the space-time directions soak up zero modes and yield an $\epsilon$-independent contribution. However, the $K3$ fermion correlator is non-trivial and exponentiates to $G_{\textrm{ferm}}(\epsilon_{+}) = \langle e^{-\check\epsilon_{+}\check J_R}\rangle$, with $\check J_R=\int(\chi^4\chi^5-\bar\chi^4\bar\chi^5)$. This is precisely the operator rotating the supercharges and is identified with the $R$-symmetry current in the field theory limit. Hence, the choice of self-dual vertices \eqref{Tvertices} correctly mimics the $\Omega$-background in the trace representation \eqref{NekrasovTrace} and is guaranteed to produce the perturbative\footnote{It turns out \cite{Antoniadis:2013mna} that it also correctly reproduces the non-perturbative part of the gauge theory partition function, as discussed in the next section.} part of the Nekrasov partition function in the field theory limit. This limit is obtained by expanding the amplitude around a Wilson line enhancement point, where both $P_L,P_R\rightarrow 0$, as a consequence of the fact that $\epsilon_\pm$ are dressed with $\langle\partial X\rangle$ and $\langle\bar\partial X\rangle$. It was shown in \cite{Antoniadis:2013bja} that the vertices \eqref{Tvertices} correspond to the self-dual field-strengths in the vector multiplet of the K\"ahler modulus $T$ of the heterotic $T^2$.


\section{Deformed Topological Amplitudes}\label{DeformedTopoAmp}

\begin{figure}[h!t]
\begin{center}
\includegraphics[width=0.75\textwidth]{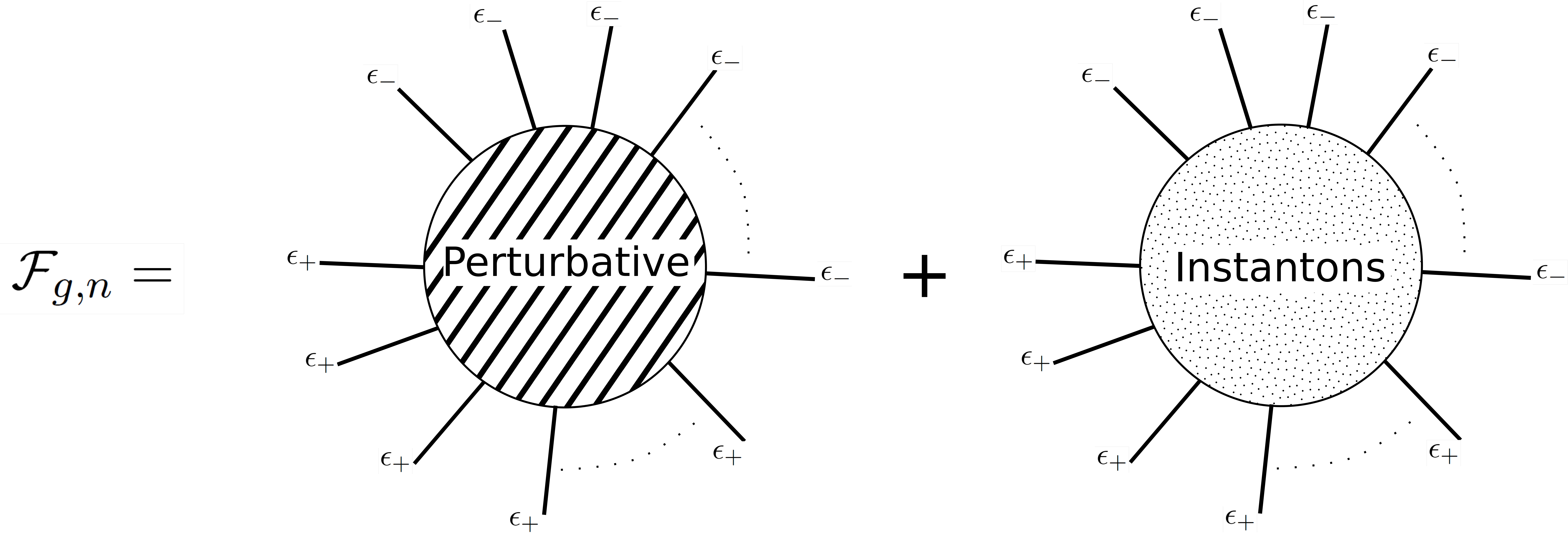}
\end{center}
\caption{The effective couplings $\mathcal F_{g,n}$ receive contributions from both perturbative and instanton sectors as presented below.}
\label{Fig-PertInst}
\end{figure}

In this section, we review the results of \cite{Antoniadis:2013bja,Antoniadis:2013mna} for the calculation of the effective couplings \eqref{GenFterms} with insertions of the self-dual $\bar T$-vectors. They receive contributions from perturbative and non-perturbative states as depicted in Fig. \ref{Fig-PertInst}. The perturbative calculation is presented in heterotic on $K3\times T^2$ in which $K3$ is realised, for convenience, as a $T^4/\mathbb{Z}_N$ orbifold, even though the results are expected to hold for a generic $K3$ compactification. On the other hand, the evaluation of the instanton effective action is shown in the dual type I setup since the latter is more suitable for realising instantons in terms of D-brane bound states \cite{Witten:1995gx, Douglas:1996uz}.

\paragraph{Perturbative results} As explained in the previous section, due to the fact that the $T^2$ coordinates $(X,\psi)$ appear holomorphically in the vertex operators, the full correlation function splits into a bosonic and a fermionic one. They can both be exponentiated and thus calculated as deformed partition functions.

First of all, the bosonic generating function takes the form
\begin{align}\label{BosGen}
G^\textrm{bos}(\epsilon_{-},\epsilon_{+})=\left< \exp\Biggr[-\tilde\epsilon_{-}\int{d^2 z~(Z^1 \bar\partial Z^2 + 
\bar{Z}^2\bar\partial\bar{Z}^1)} -  \check\epsilon_{+}\int{d^2 z~(Z^1 \partial \bar{Z}^2 + Z^2\partial\bar{Z}^1)} 
\Biggr] \right>~,
\end{align}
such that the relevant bosonic correlators contributing to $\mathcal{F}_{g,n}$ are obtained by an appropriate number of derivatives with respect to $\epsilon_\pm$. Notice that \eqref{BosGen} has the structure of the $\Omega$-deformation in space-time. It can be explicitely evaluated as a Gaussian path integral to yield \cite{Zeinassi:2014}
\begin{align}
  G^{\textrm{bos}}(\epsilon_-,\epsilon_+)&=\exp\Bigl[\zeta(2)(\tilde\epsilon_-^{\,2}\,\hat{\bar E}_2+\check\epsilon_+^{\,2}\,\hat E_2)\nonumber\\
   &+\sum_{\genfrac{}{}{0pt}{}{k\geq2}{}}\sum_{\genfrac{}{}{0pt}{}{0\leq \ell\leq k}{}}\binom{2k}{2\ell}\frac{\zeta(2k)}{k}\,\tilde\epsilon_-^{\,2\ell}\,\check\epsilon_+^{\,2(k-\ell)}\,\tau_2^{\,2(\ell-k)}\bar E(k,4\ell-2k)\Bigr]\,,\label{FullRegBos}
\end{align}
where $E(s,w)$ is the non-holomorphic Eisenstein series of weight $w$ defined as
\begin{align}
	E(s,w) \equiv \frac{1}{2}\sum_{(c,d)=1}\frac{\tau_2^{s-w/2}}{|c\tau+d|^{2s-w}}\,(c\tau+d)^{-w}\,.
\end{align}
Similarly, the fermionic correlation functions can be determined, after the spin structure sum, via the generating function\footnote{The Gaussian path integral here is to be evaluated with the fermions in the odd spin structure.} 
\begin{align}\label{FermGen}
G^{\text{ferm}}\left[\begin{matrix}
                             h \\
			     g
                            \end{matrix}
  \right](\epsilon_+)=\left\langle e^{-\epsilon_+\int (\chi^4\chi^5-\bar{\chi}^4\bar{\chi}^5)\bar{\partial}X}\right\rangle_{h,g}
\end{align}
parametrised by $\epsilon_+$, with $h$ labelling the orbifold sectors and the sum over $g$ enforcing the orbifold projections. It carries the structure of an $R$-symmetry twist as in the gauge theory partition function. The Gaussian integral yields
\begin{equation}
 G^\textrm{ferm}[^h_g](\check\epsilon_+)=\frac{\theta[^{1+h}_{1+g}](\check\epsilon_+ ;\tau)\theta[^{1-h}_{1-g}](\check\epsilon_+ ;\tau)}
{\eta^2}~ e^{\frac{\pi}{\tau_2}\check\epsilon_+^2}~.
\end{equation}
The full amplitude can then be written by including also the internal and gauge degrees of freedom \cite{Antoniadis:2013bja}. In particular, the coupling function can be obtained as the $\epsilon_-^{2g}\epsilon_+^{2n}$-term in the $\epsilon_{\pm}$-expansion of $\mathcal F^{\textrm{pert}}(\epsilon_-,\epsilon_+)$.

We now expand the full amplitude around an $SU(2)$ Wilson line enhancement point characterised by $P_L=P_R\sim W ~\rightarrow~ 0$, with $\vec W\equiv \vec Y_2-U\vec Y_1$ being the complexified Wilson line of $T^2$ along the $E_8$ directions. As expected, the field theory limit of the generating function for the refined couplings~is
\begin{align}\label{result}
\mathcal{F}^{\textrm{pert}}\left(\epsilon_-,\epsilon_+\right)&~\sim~ (\epsilon_{-}^2-\epsilon_{+}^2)\int_0^\infty\frac{dt}{t}~
\frac{-2\cos\left( 2\epsilon_+ t\right) }{\sin\left( \epsilon_--\epsilon_+\right)t ~\sin\left( \epsilon_-+
\epsilon_+\right)t} ~e^{-\mu t}
\end{align}
in terms of the BPS mass parameter $\mu\sim W$. In particular, it exhibits the correct singularity behaviour as $\mu^{2-2g-2n}$. Hence, we recover the perturbative part of the partition function of the $\mathcal{N}=2$ gauge theory in the $\Omega$-background, for an $SU(2)$ gauge group without flavours. More general gauge groups can be probed by studying different enhancement points as explained in \cite{Antoniadis:2013bja}.

\paragraph{Deformed ADHM action from string theory} We now turn to the non-perturbative part of the refined couplings $\mathcal F_{g,n}$ as displayed in the second diagram of Fig. \ref{Fig-PertInst} and review the results of \cite{Antoniadis:2013mna}. The difference with the perturbative calculation lies in the choice of the states running in the `loop'. In other words, we are interested in the instanton effective action in the presence of the background of anti-self-dual graviphotons and self-dual $\bar T$-vectors. Integrating out the instanton moduli in the latter leads to the non-perturbative contributions to $\mathcal F_{g,n}$. This extends the results of \cite{Billo:2006jm} where it was shown that the background of graviphotons calculates the unrefined ADHM partition function, \emph{i.e.} for $\epsilon_+=0$.

As explained above, we work in the type I dual of the heterotic setup. The graviphoton is mapped to itself and the $\bar T$-vectors to the $\bar S'$-vectors\footnote{Recall that in type I theory on $K3\times T^2$, $\textrm{Im}(S')$ is identified with the D5-brane coupling.}. Here, we focus on the leading order in $\alpha'$ of the non-perturbative corrections. For this, we realise Yang-Mills instantons as D5-D9 bound states with the D5-instantons wrapping $K3\times T^2$. In this context, the massless states arising from open strings with at least one endpoint lying on a D5-instanton are in one-to-one correspondence with the ADHM moduli\footnote{The D9-D9 states give rise to the $\mathcal N=2$ vector multiplet.}. Therefore, the strategy is to calculate the tree-level  couplings between the ADHM moduli and the background of graviphotons and $\bar S'$-vectors to leading order in $\alpha'$. Some of these also involve the $\mathcal{N}=2$ vector multiplet\footnote{In particular, the scalar $\tilde a$ playing the role of the Coulomb branch parameter.}. From the string theory perspective, these couplings are calculated by disc diagrams and are three- and four-point functions (higher point functions are suppressed in the field theory limit).

Compared to the heterotic theory, the main difference in the closed string sector lies in the form of the vertex operators of interest. More specifically, they have an NS-NS and a R-R part:
\begin{align}
	V^{F^G}(y,\bar y)&=\frac{1}{8\pi\,\sqrt{2}}F_{\mu\nu}^{G} \bigg[\,\psi^{\mu}\psi^{\nu}(y)e^{-\varphi(\bar y)}{\bar\Psi}(\bar y)\,+e^{-\varphi(y)}\bar\Psi(y)\psi^{\mu}\psi^{\nu}(\bar y)\nonumber\\
&-\frac{i}{2}\, e^{-\tfrac{1}{2}(\varphi(y)+\varphi(\bar y))} S_{\alpha}(y) (\sigma^{\mu\nu})^{\alpha\beta} 
S_{\beta}(\bar y)\,\epsilon^{AB}\,S_{A}(y)S_{B}(\bar y)\bigg]\,,\label{GraviphotonVertex}\\
V^{F^{\bar{S}'}}(y,\bar y)&=\frac{1}{8\pi\,\sqrt{2}}F_{\mu\nu}^{\bar{S}'} \bigg[\,\psi^{\mu}\psi^{\nu}(y)e^{-\varphi(\bar y)}{\bar\Psi}(\bar y)\,+e^{-\varphi(y)}\bar\Psi(y)\psi^{\mu}\psi^{\nu}(\bar y)\nonumber\\
&+\frac{i}{2}\, e^{-\tfrac{1}{2}(\varphi(y)+\varphi(\bar y))} S_{\dot{\alpha}}(y) (\bar{\sigma}^{\mu\nu})^{\dot{\alpha}\dot{\beta}} 
S_{\dot{\beta}}(\bar y)\,\epsilon_{\hat A\hat B}\,S^{\hat A}(y)S^{\hat B}(\bar y)\bigg]\,.\label{Tbarvector}
\end{align}
Here, $S_{\alpha}, S_{\dot\alpha}$ are space-time spin-fields whereas $S_A, S^{\hat A}$ are internal ones. It turns out that both parts of the vertex operators lead to the same coupling to the ADHM action for the $\bar S'$-vector whereas the R-R part of the graviphoton couples with an additional minus sign $(-)$ with respect to its NS-NS part, and this is due to its anti-self-duality. As dictated by the perturbative interpretation of the couplings, we identify the constant anti-self-dual field strength of the graviphotons with $\epsilon_{-}$ and the self-dual field strength of $\bar{S}'$ with $\epsilon_{+}$. Therefore, both field strenghts lead to non-trivial couplings to the ADHM action which we now recall:
\begin{align}
 S(\epsilon_-,\epsilon_+)=-\textrm{Tr}&\left\{[\chi^{\dag},a_{\alpha\dot\beta}]\left([\chi,a^{\dot\beta\alpha}]+\epsilon_{-}(a\tau_3)^{\dot\beta\alpha}\right)+\epsilon_+\left[\chi^{\dag},a_{\alpha\dot{\beta}}\right](\tau_3 a)^{\dot\beta\alpha}\right.\nonumber\\
   &\left.-\chi^{\dag}\,\bar\omega_{\dot\alpha}\left(\omega^{\dot\alpha}\chi-\tilde a\,\omega^{\dot\alpha}\right)-\left(\chi\bar\omega_{\dot\alpha}-\bar\omega_{\dot\alpha}\,\tilde a\right)\omega^{\dot\alpha}\,\chi^{\dag}-\epsilon_+\,\bar{\omega}_{\dot{\alpha}}\,{(\tau_3)^{\dot{\alpha}}}_{\dot{\beta}}\,\chi^{\dag}\,\omega^{\dot{\beta}}\right\}\,,
\end{align}
with $a$, $\chi$, $\omega$ and $\bar\omega$ being the bosonic ADHM moduli. This is precisely the $\Omega$-deformed part of the ADHM action used to calculate the (non-perturbative) Nekrasov partition function using localisation\footnote{The $\epsilon$-independent parts of the Yang-Mills and ADHM actions can be recovered in a similar fashion from the same string theory setup, see \emph{e.g.} \cite{Green:2000ke}. For a treatment without auxiliary fields, we refer to \cite{Antoniadis:2013mna,Zeinassi:2014}.}. Consequently, the instanton corrections to the refined couplings \eqref{GenFterms} involving the $\bar S'$-vectors also reproduce the non-perturbative part of the $\mathcal N=2$ gauge theory partition function in the $\Omega$-background.

It is interesting to perform the same analysis for the coupling involving the vector partner of the dilaton. Indeed, its vertex operator can be obtained from \eqref{Tbarvector} by changing the relative sign between the NS-NS and the R-R parts. Hence, all the couplings to the ADHM moduli involving the self-dual $S$-vector are trivial and cannot lead to the correct gauge theory partition function upon evaluation of the instanton path integral.


\section{Conclusions}\label{Sec:Conclusions}

We have seen that string amplitudes corresponding to the higher derivative $R^2 F_{G,(-)}^{2g-2}F_{\bar{T},(+)}^{2n}$ terms in the string effective action, expanded around an appropriate Wilson line enhancement point in moduli space, correctly reproduce the partition function of the $\mathcal{N}=2$ gauge theory $Z_{\textrm{Nek}}(\epsilon_{-},\epsilon_{+})$, perturbatively as well as non-perturbatively. These couplings provide the desired one-parameter extension of the well-known $\mathcal{N}=2$ topological couplings $F_{g,0}=F_g$, correctly realising the full $\Omega$-background.

One may eventually ask what the criterion determining this particular choice of self-dual vertices is. A partial answer is based on the arguments presented in Section \ref{Motivation}. Indeed, in order to obtain the correct perturbative Nekrasov partition function in the field theory limit, the scattering background must properly exponentiate to correctly mimic the $\Omega$-background. Recall that a very drastic constraint to our considerations was the requirement of exact solvability of our amplitude. This essentially boils down to imposing a holomorphic dependence on the $T^2$ coordinate in  choosing the self-dual vertex operators. Nevertheless, the present amplitude is only a single element in a family of amplitudes that lie in the same T-duality orbit. In fact, T-duality transformations that preserve the left-moving structure of the vertex operators also change the enhancement point. They typically lead to combinations of gauge fields involving also the $E_8$ gauge bundle, but all share the same field theory limit. We close this discussion with a conjecture. Namely, that it is unlikely for an exactly solvable (modular), truly BPS amplitude to give rise to the correct Nekrasov partition function, unless it is defined through the present non-BPS amplitudes by taking an appropriate limit in moduli space, projecting to the BPS sector. The reason is that it seems impossible to implement the fermionic $R$-symmetry twist in a modular invariant fashion, without spoiling the BPS nature of the amplitude. The same conclusion can be reached from the very nature of the higher derivative couplings \eqref{GenFterms}, \emph{i.e.} that the generalised F-terms are not super-protected against non-BPS corrections. In this sense, we conjecture that the ansatz \eqref{Tvertices} is unique, modulo the above mentioned images under T-duality.

As pointed out above, from the string perspective, the couplings we considered also receive contributions from non-BPS states and are not truly topological. On the other hand, in some appropriate decompactification limit, we expect the geometric details of the theory to be washed away and the resulting couplings, when properly mapped into the type II side using the duality dictionary, to be expressed in terms of topological invariants. This expectation is consistent with the fact that preliminary results indicate the presence of differential equations satisfied by our couplings which generalise the standard holomorphic anomaly equations of the $\mathcal F_g$'s \cite{Bershadsky:1993cx}. We believe that future work in this direction will provide interesting insights into establishing this connection.


\paragraph{Acknowledgements}

  It is a great pleasure to thank I.~Antoniadis, S.~Hohenegger and K.\,S. Narain for the enjoyable collaboration on which these results are based. We would also like to thank the organisers of the Corfu Summer Institute 2013 and the ``Workshop on Noncommutative Field Theory and Gravity" for the opportunity to present this work. This work was partly supported by the European Commission under the ERC Advanced Grant 226371.



\bibliographystyle{unsrt}

\begin{thebibliography}{99}


\bibitem{Antoniadis:2013bja} 
  I.~Antoniadis, I.~Florakis, S.~Hohenegger, K.~S.~Narain and A.~Zein~Assi,
  Nucl.\ Phys.\ B {\bf 875} (2013) 101
  [arXiv:1302.6993 [hep-th]].

\bibitem{Antoniadis:2013mna}
  I.~Antoniadis, I.~Florakis, S.~Hohenegger, K.~S.~Narain and A.~Zein Assi,
  Nucl.\ Phys.\ B {\bf 880} (2014) 87
  [arXiv:1309.6688 [hep-th]].

\bibitem{Witten:1988xj}
  E.~Witten,
  Commun.\ Math.\ Phys.\  {\bf 118} (1988) 411.

\bibitem{Antoniadis:1996qg}
 I.~Antoniadis, E.~Gava, K.~S.~Narain and T.~R.~Taylor,
 Nucl.\ Phys.\ B {\bf 476} (1996) 133
[hep-th/9604077].

\bibitem{Antoniadis:1993ze} I.~Antoniadis, E.~Gava, K.~S.~Narain and T.~R.~Taylor,
 Nucl.\ Phys.\ B {\bf 413} (1994) 162 [hep-th/9307158].


\bibitem{Gopakumar:1998ii} R.~Gopakumar and C.~Vafa,
 hep-th/9809187, hep-th/9812127.

\bibitem{Losev:1997bz}
  A.~Losev, N.~Nekrasov and S.~L.~Shatashvili,
  In *Cargese 1997, Strings, branes and dualities* 359-372
  [hep-th/9801061]. 

\bibitem{Moore:1997dj}
  G.~W.~Moore, N.~Nekrasov and S.~Shatashvili,
  Commun.\ Math.\ Phys.\  {\bf 209} (2000) 97
  [hep-th/9712241].

\bibitem{Dijkgraaf:2006um}
  R.~Dijkgraaf, C.~Vafa and E.~Verlinde,
  [hep-th/0602087].

\bibitem{Awata:2005fa}
  H.~Awata and H.~Kanno,
  JHEP {\bf 0505} (2005) 039
  [hep-th/0502061].

\bibitem{Iqbal:2007ii} A.~Iqbal, C.~Kozcaz and C.~Vafa,
 JHEP {\bf 0910} (2009) 069 [hep-th/0701156].

\bibitem{Zeinassi:2014}
 A. Zein Assi,
 Ph.D. Dissertation, Ecole Polytechnique, TEL-00942993,
  [arXiv:1402.2428 [hep-th]].

\bibitem{Nekrasov:2002qd} N.~A.~Nekrasov,
 Adv.\ Theor.\ Math.\ Phys.\  {\bf 7} (2004) 831 [hep-th/0206161].

\bibitem{Florakis:2010ty}
  I.~Florakis, C.~Kounnas and N.~Toumbas,
  Nucl.\ Phys.\ B {\bf 834} (2010) 273
  [arXiv:1002.2427 [hep-th]].

\bibitem{Morales:1996bp} 
  J.~F.~Morales and M.~Serone,
  Nucl.\ Phys.\ B {\bf 481}, 389 (1996)
  [hep-th/9607193].

\bibitem{Antoniadis:2010iq}
 I.~Antoniadis, S.~Hohenegger, K.~S.~Narain and T.~R.~Taylor,
  Nucl.\ Phys.\ B {\bf 838} (2010) 253
  [arXiv:1003.2832 [hep-th]].

\bibitem{Witten:1995gx}
  E.~Witten,
  Nucl.\ Phys.\ B {\bf 460} (1996) 541
  [hep-th/9511030].

\bibitem{Douglas:1996uz}
  M.~R.~Douglas,
  J.\ Geom.\ Phys.\  {\bf 28} (1998) 255
  [hep-th/9604198].

\bibitem{Billo:2006jm}
  M.~Billo, M.~Frau, F.~Fucito and A.~Lerda,
  JHEP {\bf 0611} (2006) 012
  [hep-th/0606013].

\bibitem{Green:2000ke}
  M.~B.~Green and M.~Gutperle,
  JHEP {\bf 0002} (2000) 014
  [hep-th/0002011].

\bibitem{Bershadsky:1993cx}
  M.~Bershadsky, S.~Cecotti, H.~Ooguri and C.~Vafa,
  Commun.\ Math.\ Phys.\  {\bf 165} (1994) 311.





\end{thebibliography}

\vfill\eject

\end{document}